\documentclass[twocolumn,noshowpacs,preprintnumbers,amsmath,amssymb]{revtex4}
\usepackage[version=3]{mhchem} % Formula subscripts using \ce{}

\usepackage{epsfig}
\usepackage{graphicx}% Include figure files
\usepackage{dcolumn}% Align table columns on decimal point
\usepackage{epic,eepic}
\usepackage[usenames]{color}
\usepackage{rotating}
\usepackage{amsmath}
\usepackage{amssymb}
\usepackage{chemfig}
\newcommand\setpolymerdelim[2]{\def\delimleft{#1}\def\delimright{#2}}
\def\makebraces[#1,#2]#3#4#5{%
\edef\delimhalfdim{\the\dimexpr(#1+#2)/2}%
\edef\delimvshift{\the\dimexpr(#1-#2)/2}%
\chemmove{%
\node[at=(#4),yshift=(\delimvshift)]
{$\left\delimleft\vrule height\delimhalfdim depth\delimhalfdim width0pt\right.$};%
\node[at=(#5),yshift=(\delimvshift)]
{$\left.\vrule height\delimhalfdim depth\delimhalfdim width0pt\right\delimright_{\rlap{$\scriptstyle#3$}}$};}}
\setpolymerdelim()
\newcolumntype{e}[1]{D{.}{.}{#1}}
\begin{document}

%\begin{frontmatter}

\title{Simulations of water and hydrophobic hydration using a neural network potential}
\author{Alexander S. Lyons, Jr. and Steven W. Rick}
\email{srick@uno.edu}
\affiliation{
Department of Chemistry, University of New Orleans,
 New Orleans, LA, 70148
 }

\begin{abstract}

Using a neural network potential (ANI-1ccx) generated from quantum data on a large data set of molecules and pairs of molecules, 
isothermal, constant volume simulations demonstrate that the model can be as accurate as {\it ab initio} molecular dynamics for simulations of pure liquid water and
the aqueous solvation of a methane molecule.
No theoretical or experimental data for the liquid phase is used to train the model,
suggesting that the ANI-1ccx approach is an effective method to link high level {\it ab initio} methods to potentials for large scale simulations.

\end{abstract}

\maketitle 

%\end{frontmatter}

 \section{Introduction}
 
Conventional molecular simulations require
 a model, or force field, which can involve 
 developing potential functions that capture the essential physics of the system and 
 optimizing a large set of parameters.
 The parameterization often uses a combination of quantum calculations and experimental data. 
 The resulting models can in many cases be accurate and efficient, but the time required in their development and the need for experimental data 
 can be limiting, especially for the simulation of new materials.
 Simulations without models can be carried out using {\it ab initio} molecular dynamics (AIMD),\cite{CarParrinello,PaquetViktor}
 however these are limited in system size and time scales and can require using relatively low level density functional theory (DFT) methods. 
 
 Machine learning (ML) methods offer a means to carry out simulations with high accuracy without a potential function or experimental 
 data.\cite{Doren1995,Manzhos,BehlerParrinello,Knoll2015,DeVita2015,Behler2016,SmithRoitberg2018,ANI-1CCX}
 These methods  use {\it ab initio} calculations to train a neural network representation of a potential energy surface. 
 The goal of these methods is to develop potentials that can be used just as {\it ab initio} methods, so they can treat new materials and
 reacting systems, but are much faster than AIMD. 
 Central to these methods are descriptors, or generalized coordinates,
 that transfer the high dimensional set of coordinates of the system to a set of inputs, the descriptors, for the ML algorithms.
 Several descriptors have been developed that satisfy the necessary physical conditions
 of being invariant with respect to translation and rotation, as well as exchange of identical atoms.\cite{Doren1995,BehlerParrinello,Bartok,Knoll2015,DeVita2015,SmithIsayevRoitberg,ChmielaTkatchenko,WillattMusilCeriotti}
 Other approaches, which do not build in these symmetries, have been successful as well.\cite{Liu2019}
 For molecular dynamics simulations, the descriptors
 should also be continuous and differentiable so they can be used to calculate forces.\cite{BehlerReview}
Using these descriptors,  artificial neural networks (NN) with multiple layers  learn to reproduce {\it ab initio} energies 
for a selected\cite{BehlerReview,SmithIsayevRoitberg} reference set of molecules.
While the resulting models are one to two orders of magnitude slower than conventional force fields, they 
can  be several orders of magnitude faster to compute than DFT methods.\cite{BehlerReview} 

One of the most promising features of ML in general is their applicability to systems for which they have not been trained.
In this work,  we will use a  NN potential to simulate liquid water and a methane molecule solvated by water. 
Water and hydrophobic solvation were chosen for their significance and the abundance of experimental data
for comparison.
NN potentials for water have been developed and have shown high accuracy in reproducing the properties of liquid water and ice
over a range of state points.\cite{BehlerPNAS,NatarajanBehler,WangYang2018,Singraber,ChengCeriotti,MorawietzSharmaBehler}
These models were trained using water clusters or liquid water, specifically for water. Some use a 
purely machine-learning approach\cite{BehlerPNAS,NatarajanBehler,Singraber,ChengCeriotti} while
others add combine ML with 
physical potentials, such as Coulombic interactions.\cite{WangYang2018,MorawietzSharmaBehler} 
Other approaches use ML to optimize parameters for a classical force field.\cite{LiLiPackard,Beraau}
These and other ML methods for molecular simulations,\cite{Manzhos,Bartok,DeVita2015,ChmielaTkatchenko,WillattMusilCeriotti}  have
recently been compared.\cite{NguyenPaesani}. 
The point of the present work is not to develop or compare water models, but to test a general purpose NN 
potential.
Can a model not trained for water, or the condensed phases, be effective for liquid water?
The NN potential of Roitberg and co-workers.\cite{SmithIsayevRoitberg,SmithRoitberg2018,ANI-1CCX} was chosen for its generality and ease of use.
This NN potential, ANI-1ccx,\cite{ANI-1CCX}  was trained 
 for molecules made up of H, C, N, and O atoms
on a large, diverse data set  to reproduce 
energies at the coupled cluster level.
The training  dataset contains structures of single molecules and dimers only.\cite{SmithRoitberg2018} 
We use the model for simulations in the liquid phase, as a test of the transferability of the model to the condensed phase.

\section{Methods}
\noindent
 {\bf The ANI-1ccx potential.} Full details of the model have been published by Roitberg and co-workers;\cite{SmithRoitberg2018} 
 here we aim to provide a summary of the main ideas. This method builds on the NN potential of Behler and Parrinello.\cite{BehlerParrinello}
 The total energy of the system, E, is given by a sum over a contribution from each atom, $i$,
 \begin{equation}
 E=\sum_{i} E_i
 \label{eq:e1}
 \end{equation}
 where E$_i$ is the atomic contribution to the energy.
 The NN is trained to reproduce E$_i$  for each atom, with a different NN for each element. Each element is treated by the same set of NN,
 independent of its chemical environment. 
 ANI-1ccx has one NN for each element with five layers, with 3 hidden layers between the input and output layers, each with a different number of nodes.
 The NN takes as input descriptors, termed atom centered symmetry functions\cite{BehlerParrinello} or atomic environment variables\cite{SmithIsayevRoitberg}
 which are determined from the three dimensional structure.
 There are two types of symmetry functions, radial and angular.  The radial functions are given by
 \begin{equation}
 G_i^R = \sum_{j \ne i} e^{-\eta (r_{ij} - R_s )^2 } f_c(r_{ij}) 
 \label{eq:gr}
 \end{equation}
 where $r_{ij}$ is the distance between atoms i and j, $\eta$ and R$_s$ are fixed parameters for a particular node, and $f_c$ is a cut-off function that smoothly
 goes to zero at a length R$_C$. In the ANI-1ccx potential, the $\eta$ parameter is fixed and multiple R$_S$ values are used, equally spaced
 out to the cut-off  length of 5.2 {\AA}.
 
 The angular function is
 %\begin{equation}
 \begin{multline}
 G_i^A =2^{1-\zeta} \sum_{j,k \ne i} (1+\cos(\theta_{ijk} - \theta_S ) )^{\zeta} \\
 \exp\left[ -\eta \left((r_{ij}+r_{ik})/2 - R_s \right)^2 \right] f_c(r_{ij}) f_c(r_{ik}) 
 \label{eq:ga}
 %\end{equation}
 \end{multline}
 where $\theta_{ijk}$ is the angle between atoms $i$, $j$, and $k$, with $i$ at the vertex. The parameters $\theta_S$ and $R_S$ center the function
 at a particular angle and distance, with $\zeta$ and $\eta$ determining the widths of the functions. 
 This form is different from that of Behler and Parrinello and allows for greater spacial and angular resolution of an atom's environment.
 The angular function establishes the angular potentials between both bonded and non-bonded atoms, allowing for 
 many-body polarization to be treated at the three-body level. 
 
 It is worth emphasizing that these potentials are not using ML to optimize a potential model, although that can be done.\cite{Beraau}.
 Equations \ref{eq:gr} and \ref{eq:ga} are fairly agnostic functions, acting as 
 windows to represent an atom's environment. The two main approximations that go into the models are 
 that they do not go beyond a cut-off length of R$_C$ and non-additive effects are only included at the three-body level.
  
 The ANI-1ccx model was trained using 5.2 million conformations of 64 thousand molecules containing C,N,O, and H atoms with energies
 evaluated at the DFT level. Of those, 500 thousand were selected 
 for retraining  using transfer learning\cite{TaylorStone}.
The retraining was done using energies calculated approaching the accuracy of the coupled cluster with full treatment of singles and doubles, and triples calculated with perturbation 
theory extrapolated to the complete basis let limit (CCSD (T)/CBS),\cite{ANI-1CCX} using the 
domain localized, local pair natural orbital DPLNO-CCSD(T) method.\cite{Neese}
The conformations included dimers, energy minimized structures, and structures generated from normal mode sampling to generate structures
away from energy minima.\cite{SmithIsayevRoitberg,SmithRoitberg2018}

\noindent
{\bf Simulation details.}  Simulations were done using the Atomic Simulation Environment (ASE) package.\cite{Hjorth_Larsen_2017} 
The ANI-1ccx potential implemented in ASE was downloaded from GitHub 
(https:$\slash\slash$github.com$\slash$isayev$\slash$ASE$\_$ANI).
The liquid water simulations used 128 molecules and the water/methane simulations used 128 water plus one methane molecule.
Simulations were run in the NVT ensemble at a density of 1.0 g cm$^{-3}$, using Langevin thermostatting, with a time step of 0.1 fs, and at a temperature of 300 K.
Initial structures were taken from equilibrated simulations using the SPC/E model.\cite{SPCE}
The simulations were equilibrated for 5 ps and data was collected over an additional 10 ps for water. The methane/water simulations collected data over 10 ps.
The simulation times are similar to those used in AIMD simulations of water, which can range from 5 to 50 ps.\cite{MaZhangTuckerman,Willow2015,PastanaAIMD2017,WuAIMD2018,LiuHeZhangQi,Chen10846}
In our hands, with a small 4-node GPU-enabled (NVIDIA K-80; Intel Haswell) compute cluster, 
1 ps of simulation time takes about 5 days. 
The relatively long wall clock times are not due to inherent computational cost of the ANI-1ccx potential, but in its implementation.
The ASE program, written in python, integrates well with the ANI ML potential, but it is not as fast as other simulation packages. 

The dynamical properties, the diffusion constant and the Nuclear Magnetic Resonance (NMR) relaxation time, $\tau_{NMR}$, were
calculated using 6  NVE simulations of 2 ps each.
The diffusion constant is calculated form the mean-square displacement of a molecule's center of mass using  the Einstein relation, fit to the range 
 from 0.5 to 1.2 ps of each 2.0 ps trajectory. Corrections for the finite size of the system were made for the diffusion constant, according to
  D=D(L) + 2.837 k$_B$ T/(6$\pi\eta$L), where D(L) is the diffusion constant for simulation with side length L, k$_B$ is Boltzmann's constant, and $\eta$ is the
  viscosity.\cite{YehHummer} We use the experimental value for the viscosity of 8.9$\times$10$^{-4}$ kg m$^{-1}$ s$^{-1}$.\cite{HarrisWoolf}
  The NMR relaxation time, $\tau_{NMR}$, was found from the second order Legendre polynomial of the time correlation function for rotations around the 
  axis connecting the hydrogen atoms.\cite{Impey} 
  We fit this to an exponential function over the range from 1.0 to 1.5 ps.
  
Voronoi tessellation was used to estimate the volume occupied by solvent and solute molecules.
The Voronoi method of assigning a region of space to a
specific atom assumes that all atoms are of equal size. 
Alternative Voronoi volumes  can be defined which take into account molecular size by assigning a molecule a radius, r$_A$.\cite{AnishchikMedvedev,VoloshinGeiger}
In the surface-based (S-cell) method, a region of space is assigned to an atom using the distance to the surface of a sphere of radius R$_A$ around the atom. 
Radial distribution functions were used to assign r$_A$.
The water radial distribution function (RDF) reaches one at a distance of about 2.6 {\AA}, giving a water r$_A$ of 1.3 {\AA}.
The water-methane RDF reaches one at a distance of 3.5 {\AA}, giving a methane r$_A$ of 2.2 {\AA}, after subtracting the water radius.
Monte Carlo integration was used to calculate both the normal  and the S-cell Voronoi volumes using 10$^6$ random points per configuration.

\section{Results}

\begin{figure}[h]
    \centering 
        \includegraphics[bb=0.0cm 0.0cm 7.08cm 7.1cm]{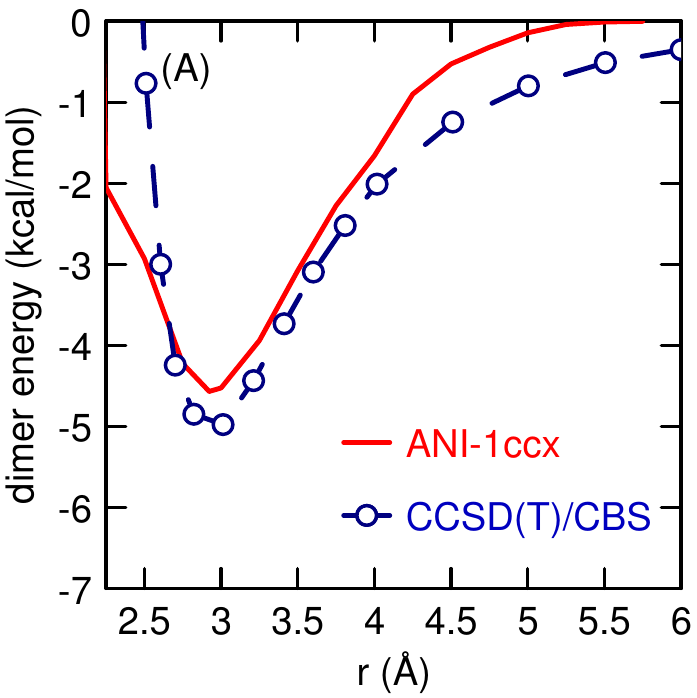}
        \includegraphics[bb=0.0cm 0.0cm 7.08cm 7.1cm]{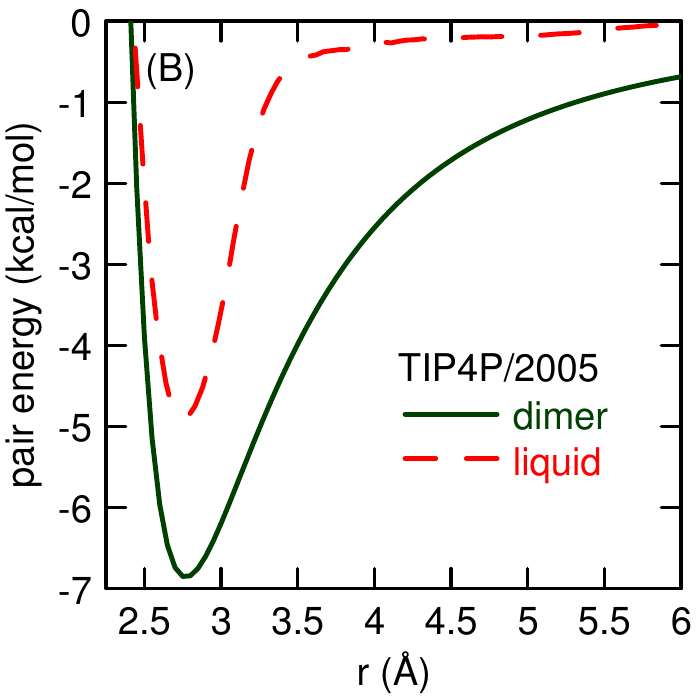}
        \caption{(A) Minimized energy of the water pair as a function of oxygen separation comparing the ANI-1ccx model to {\it ab initio} results\cite{Metz}. (B) Minimized dimer energy (solid green line) and average pair energy for the liquid at 300 K (dashed red line)  for the TIP4P/2005 model.\cite{AbascalVega} }.
                 \label{fig:dimerE}
\end{figure}

\noindent
{\bf Water.} The minimized water monomer has a OH bond length of 0.958 {\AA} and a bond angle of 104.6$^{\circ}$, 
very close to the experimental values of  0.957 {\AA}  and 104.474$^{\circ}$,\cite{Franks} and an energy of -2078.50 eV. At 300 K, the average monomer energy is -2078.43 eV.
The water dimer energies for the ANI-1ccx model are compared to the {\it ab initio} results, at the CCSD(T) level in the CBS limit,\cite{Metz}
 as shown in Figure \ref{fig:dimerE}(A).\cite{GuidezGordon}
Note that the range of the ANI-1ccx model is shorter than the {\it ab initio} results, going to zero above 5.2 {\AA},
the cut-off of the potential.
Results from liquid phase simulations using pair potentials assess the importance of the long-ranged interactions.
 Figure \ref{fig:dimerE}(B) shows the dimer energy for TIP4P/2005 model\cite{AbascalVega} and like other non-polarizable models it has a deeper minimum, since it
 is optimized to treat liquid phase interactions. 
 Since the TIP4P/2005 model is pair-wise additive, the interaction energies between pairs in the liquid phase can be determined. These
 results, shown by the red dashed line in Figure \ref{fig:dimerE}(B), show that beyond the nearest-neighbor distances the interaction
 between pairs is near zero. At distances beyond 3.5 {\AA}, the geometries between pairs are not optimal---liquid structure is largely 
 determined by optimizing nearest neighbors interactions---so the average interactions are small. 
 The longer-ranged interactions are not as  significant as the dimer energies suggest.

 The water trimer was used to assess the size of non-additive contributions to the energy. For the minimum energy cyclic structure, with three hydrogen 
 bonds, the ANI-1ccx potential
 gives a binding energy (the energy of the timer minus three times the energy of the monomer) of -15.64 kcal/mol, or -5.2 kcal/mol per molecule.
\begin{table}
\renewcommand{\arraystretch}{1.0}
\setlength{\tabcolsep}{5pt}
\caption{Cluster binding energies (in kcal/mol) for the ANI-1ccx model, CCSD(T)/aug-cc-pVDZ\cite{Miliordos},
and TIP4P/2005\cite{ClusterTIP4P}.}
\vspace{1. mm}
\begin{tabular}{llll}
\hline
\hline
N &  ANI-1ccx  & CCSD(T) & TIP4P/2005 \\
\hline
2 & -4.57 & -5.29 & -6.81 \\
3 & -15.64 & -16.33 & -18.40 \\
4 & -27.47 & -28.54 & -30.66 \\
5 & -36.22 & -37.43 & -39.99 \\
6 prism & -46.06 & -48.82 & -51.03 \\
6 cage & -45.87 & -48.38 & -52.11 \\
6 book & -45.87 & -47.71 & -50.78 \\
6 cyclic & -45.41 & -46.17 & -48.82 \\
 \hline
 \hline
 \end{tabular}
\label{tab:clusters}
\end{table}
 %High level {\it ab initio} estimates give trimer minimum energies  around -5.5 kcal/mol.\cite{KeutschCruzanSaykally,Miliordos}
 The dimer minimum is -4.57 kcal/mol, so the trimer forms hydrogen bonds $-$0.64 kcal/mol  or 12\% stronger than the dimer.
 Various {\it ab initio} methods find that the three-body contribution to the energy is around 11\% to 14\%.\cite{KeutschCruzanSaykally}
 The minimum energies for clusters of up to six molecules are given in Table {\ref{tab:clusters}. For the N=6 cluster, energies for 4 local minima structures\cite{Miliordos}
 are also given. These are compared to {\it ab initio} results at the CCSD(T)/aug-cc-pVDZ level\cite{Miliordos} and the TIP4P/2005 model, using
 previously published results for N=2-5\cite{ClusterTIP4P} and our own calculations for the hexamer structures. 
 The  CCSD(T)/aug-cc-pVDZ results give a three-body contribution to the hydrogen bond energy ($\delta$E=(E(N=3) $-$ 3 E(N=2))/3) of $-$0.15 kcal/mol, so less than the
 three-body terms from the ANI-1ccx potential. 
 For the non-polarizable TIP4P/2005 model, $\delta$E is +0.68 kcal/mol. It cannot be negative, without many-body terms, and the positive value
 reflects that in the trimer, the rigid TIP4P/2005 model cannot form completely optimal hydrogen bonds between each pair.
 For the larger clusters, the agreement between  ANI-1ccx and the CCSD(T) calculations is pretty good, with the NN model consistently giving higher energies. It correctly ranks the prism structure as the lowest energy hexamer structure, although it finds the cage and book structure to have equal energies, rather than the cage being lower.
 The TIP4P/2005 model, like other empirical liquid state models,\cite{JamesWales} does not correctly give the lowest energy hexamer structure.
 
The properties of liquid water as predicted by the ANI-1ccx potential are given in Table {\ref{tab:thermo}. The enthalpy of vaporization is
calculated from $\Delta$H$_{vap}$=$\langle$E$\rangle_{gas}$ $- \langle$ E $\rangle_{liquid}$/N +RT, 
where $\langle$E$\rangle_{gas}$ is the energy of an isolated gas-phase molecule, $\langle$E$\rangle_{liquid}$ is the average energy of the liquid with N molecules
and R is the ideal gas constant. It is in pretty good agreement with the experimental value of 10.5 kcal/mol,\cite{Dorsey} although
overestimated by about one kcal/mol.
\begin{table}
\renewcommand{\arraystretch}{1.0}
\setlength{\tabcolsep}{5pt}
\caption{Properties of liquid water a temperature of 300 K and a density of 1 g cm$^{-3}$. \hspace{35. mm}} 
\vspace{1. mm}
\begin{tabular}{lllll}
\hline
\hline
& $\Delta$H$_{vap}$ & P & D & $\tau_{NMR}$ \\
 & (kcal/mol) & (kbar) & (10$^{-9}$ m$^2$ s$^{-1}$) & (ps) \\
 \hline
 ANI-1ccx & 11.8$\pm$0.5 & -1.7$\pm$0.3 & 2.0$\pm$0.3 & 3.1$\pm$0.8 \\
 experiment & 10.5$^a$ &  0.001 &  2.30$^b$ & 2.46$^c$ \\ 
  \hline
 \hline
 \end{tabular}
\label{tab:thermo}
%\end{table}
\hspace{-10. mm}
a. Reference \cite{Dorsey}
b. Reference \cite{Krynicki}
c. Reference \cite{Jonas} 
\end{table} 
The pressure is negative, which, along with the overestimated $\Delta$H$_{vap}$, indicates that the balance between attractive and repulsive interactions is shifted towards attractive interactions.
AIMD with simple generalized gradient approximation (GGA) functionals tend to underestimate the density,
consistent with an overestimated pressure at the experimental density,\cite{Schmidt,GaidukGalli} 
The addition of dispersion interactions leads to improvement.\cite{Schmidt,GaidukGalli,MaZhangTuckerman,BehlerPNAS,Willow2016}
More recent simulations with meta-GGA functionals, which have better exchange interactions, give larger densities, above the 
experimental value.\cite{Chen10846,PastanaAIMD2017}
The ANI-1ccx potential gives good agreement for both dynamical properties, the translational diffusion constant and the rotatational constant, $\tau_{NMR}$, 
compared to experiment.\cite{Krynicki,Jonas}

Given that the potential has no charges and no electrostatic moments of any kind, the model does not have a dielectric response. The dielectric constant
will be one. By enlarging the training set to include ions, a NN model could be trained to respond to the field of an ion. This might require
extending the range of the potential, although to fully capture the dielectric response of water, methods that explicitly treat long-ranged interactions
may be necessary.\cite{GrisafiPRL,GrisafiCeriotti}

 The radial distribution functions for water are shown in Figure \ref{fig:grs}. The oxygen-oxygen,\cite{Skinnerrdf2014} 
 oxygen-hydrogen,\cite{Soper2000} and hydrogen-hydrogen\cite{Soper2000} experimental results are also shown.
 Also shown for comparison are the AIMD results, at CCD/aug-cc-pVDZ level, of Liu, {\it et al.}, which give very good agreement with experiment.\cite{LiuHeZhangQi}
 The agreement for the ANI-1ccx model, particularly for the oxygen-oxygen g(r), is very good for both peak positions and heights. The model does give a small
 probability for pairs at low r values. This region below r$< $2.5 {\AA} integrates to 0.07, indicating that about 1 out of every 14 molecules (1/0.07) 
 has a neighbor at these close distances. 
 \begin{figure}[h]
    \centering 
        \includegraphics[bb=0.0cm 0.0cm 8.14cm 15.87cm]{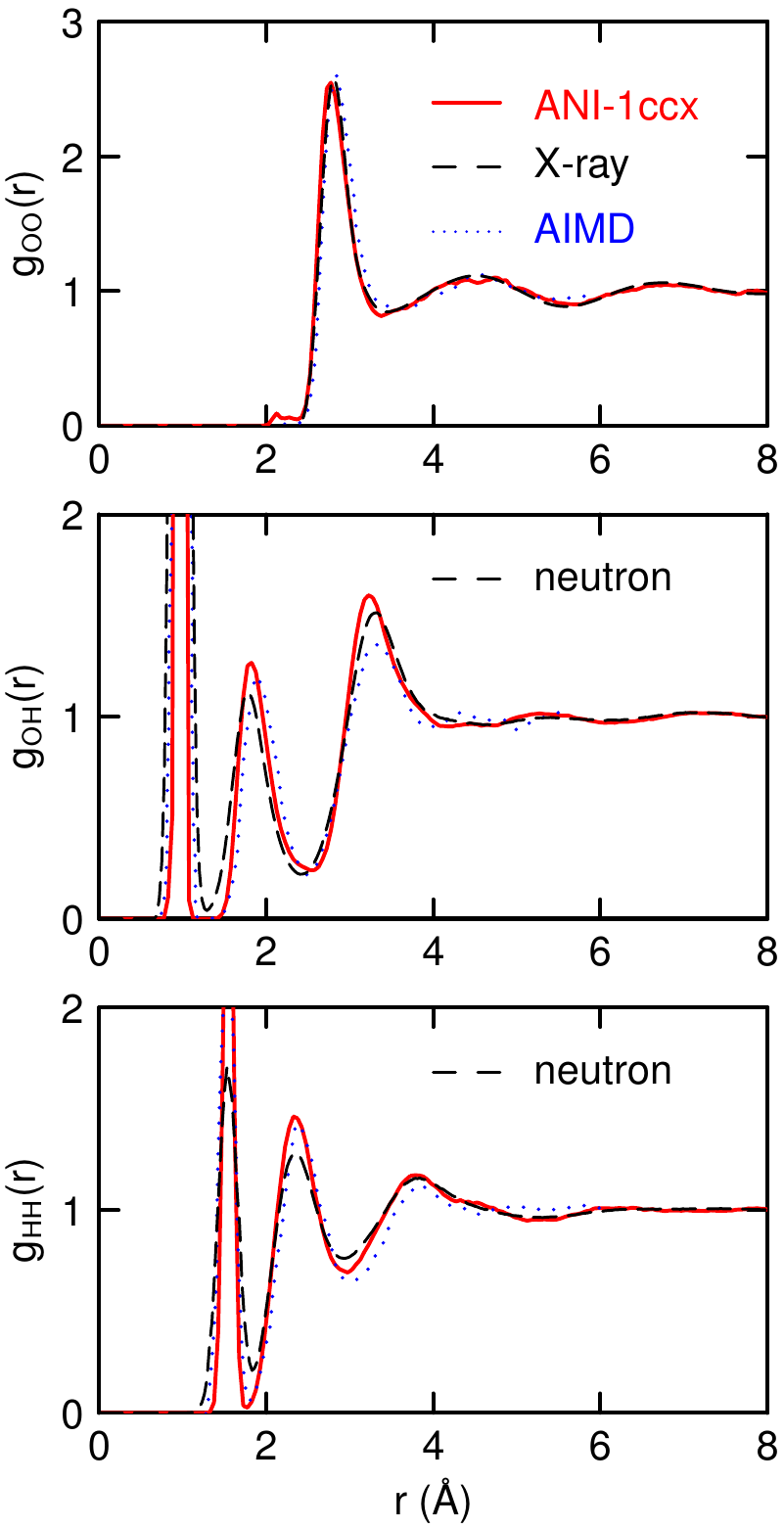}
\caption{Water-water radial distribution functions for the  ANI-1ccx model (red lines), AIMD\cite{LiuHeZhangQi} (blue dotted lines), X-ray\cite{Skinnerrdf2014} and  neutron diffraction data\cite{Soper2000,Soper2013} (black dashed lines).}
\label{fig:grs}
\end{figure}
 This feature appears a bit noisy in the g(r) since these close pairs are relatively uncommon. In these structures,
 the hydrogen between the two oxygen atoms is shifted away from the line connecting the two oxygens, so that there is not a linear hydrogen bond. Other than the
 oxygen pairs being very close, there is no atypical structures, like three water rings, formed. 
 The oxygen-hydrogen distance does not get short in this structure, so the radial distribution functions involving hydrogens do not show any unusual
 features.
 The close structures result from an oxygen-oxygen potential that is too soft at small distances (Figure \ref{fig:dimerE}(A)) and,
 like the pressure results, indicates that the repulsive interactions are underestimated.
 The potential is only too soft at distances below 2.6 {\AA}, so the oxygen-oxygen radial distribution function for most of the first peak, at distances
 larger than  2.6 {\AA}, is unaffected. 
 For the oxygen-hydrogen and hydrogen-hydrogen radial distribution functions, the peaks are a little too structured, as are some of the results from
 AIMD.\cite{Forster-Tonigold,Chen10846,MaZhangTuckerman}
 Other ML potentials have given very accurate radial distribution functions for water as well.\cite{NatarajanBehler,ChengCeriotti,WangYang2018}
 
 Nuclear quantum effects (NQE) from the quantum nature of the hydrogen atom lead to measurable structural and thermodynamic
 differences between light and heavy water.\cite{Soper2013,Stefan}
 In running classical dynamics using {\it ab initio}-derived potentials, these effects are neglected.
 Simulations using path integrals to include quantum effects find that the inclusion of quantum effects decreases the heights and broadens
 with widths of the O-H and H-H RDFs,\cite{SternBerne,SpureKuhne,ChengCeriotti} but has a smaller effect on the O-O RDF.\cite{SternBerne,MarsalekMarkland,ChengCeriotti}
Consistent with these results,  neutron diffraction results for light and heavy water find that heavy water is more structured that light
water.\cite{SoperBenmore}
The negative pressure of the  ANI-1ccx potential may be partially due the classical treatment of hydrogens as well.
NQEs have been shown to increase the density of water,\cite{ChengCeriotti,SpureKuhne,FanourgakisXantheas2008} 
and therefore increase the pressure.
Other studies suggest that quantum effects decrease the density.\cite{FanourgakisXantheas2006,PaesaniIuchiVoth}
Experimentally, heavy water has smaller density (1.1044 g cm$^{-3}$\cite{NAKAMURA1995127} at 25$^{\circ}$) than would be predicted using  the
density of light water (0.9971  g cm$^{-3}$\cite{Kell}) and multiplying by the mass ratio (20/18), to give 1.1079 g cm$^{-3}$, implying that NQEs 
increase the density.
Path integral simulations could be run with the ANI-1ccx potentials, but keeping with more typical, and less costly, simulation methods, purely classical
simulations were used.
  
Hydrogen bonds are defined using the geometric definition of Luzar and Chandler,\cite{LuzarChandler} in which the oxygen-oxygen distance
is less that 3.5 {\AA} and the angle, $\beta$, is less than 30$^{\circ}$, where $\beta$ is the angle between the vector connecting to the hydrogen bond donating oxygen and the hydrogen bond accepting oxygen and the vector connecting the donating oxygen to the hydrogen involved in the hydrogen bond.
The model gives an average number of hydrogen bonds of 3.65$\pm$0.07 (Table \ref{tab:structure}). 
\begin{table}[b] 
\renewcommand{\arraystretch}{1.0}
\setlength{\tabcolsep}{5pt}
\caption{The average number of hydrogen bonds and the average tetrahedral order parameter for liquid water at 300 K and 1 g cm$^{-3}$.}
\vspace{1. mm}
\begin{tabular}{lll}
\hline
\hline
& hydrogen bonds & Q \\
\hline
 ANI-1ccx & 3.65$\pm$0.07 & 0.66$\pm$0.02 \\
 Neutron/EPMC & 3.58$^a$ & 0.562$^b$ \\
 AIMD & 3.60$-$3.85$^c$ & 0.7$-$0.9$^d$ \\
 TIP4P/2005 & 3.66  & 0.668 \\
  \hline
 \hline
 \end{tabular}
 
\label{tab:structure}
a. Reference \cite{SoperBruniRicci} 
b. Reference \cite{SoperBenmore} 
c. Reference \cite{WuAIMD2018} \\
d. References \cite{Chen10846,WuAIMD2018} 
\end{table}
Estimates from empirical potential Monte Carlo (EPMC) using neutron diffraction
data\cite{SoperBruniRicci} gives an average hydrogen bond number of 3.58 and AIMD with various DFT methods give
hydrogen bond numbers of 3.60 to 3.85,\cite{WuAIMD2018} all using the same hydrogen bond definition as used here.
The results for the distribution of $\beta$ for neighbors within the first solvation shell, at a distance less than the minimum of the oxygen-oxygen pair
correlation function (3.4 {\AA}), are shown in Figure \ref{fig:pbeta}(A).
\begin{figure}[h]
    \centering 
        \includegraphics[bb=0.0cm 0.0cm 7.47cm 7.09cm]{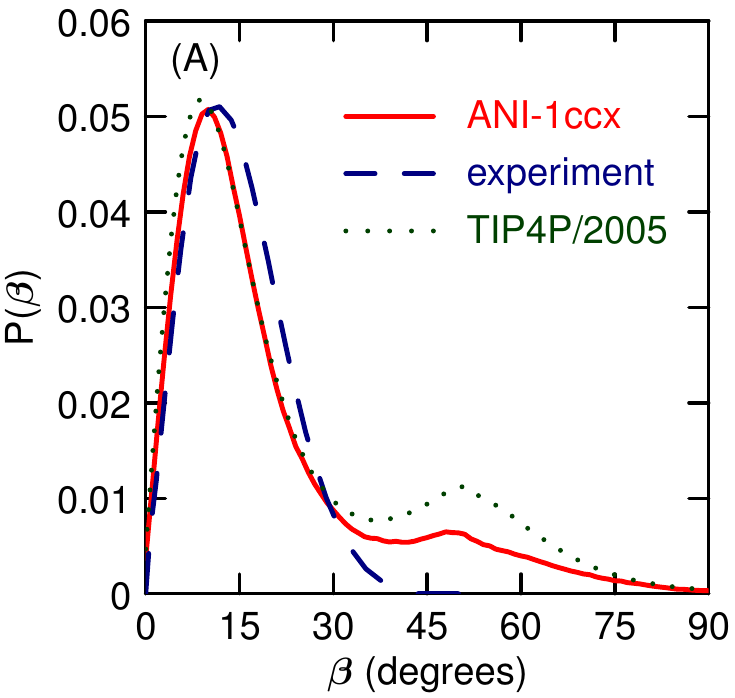}
        \includegraphics[bb=0.0cm 0.0cm 7.09cm 7.09cm]{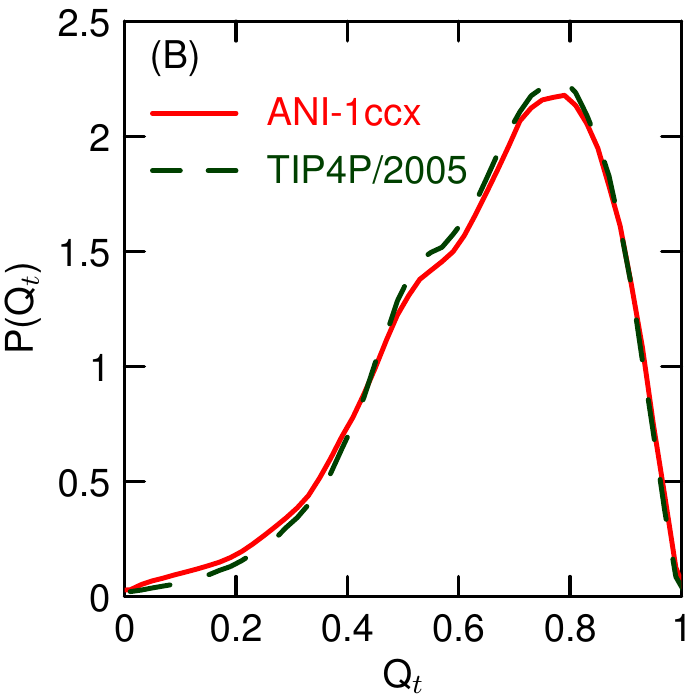}
\caption{(A) Distribution of the hydrogen bond angles for liquid water from the ANI-1ccx model (red line), the {\it ab initio}/experimental result\cite{PhysRevLett.90.075502} (blue dashed line) and the TIP4P/2005 model (green dotted line). (B) Distribution of the tetrahedral order parameter for liquid water from the ANI-1ccx model (red line) and the TIP4P/2005 model (green dashed line).}
\label{fig:pbeta}
\end{figure}
Combining measurements of the proton magnetic shielding tensor with DFT calulations, Modig, Pfrommer, and Halle constructed 
distribution functions of the hydrogen bond angle, $\beta$.\cite{PhysRevLett.90.075502} 
Our results are in pretty close
agreement with those results,  except for a tail of the distribution at angles beyond 40$^{\circ}$, indicating nearest-neighbors
which are not hydrogen bonded.
For TIP4P/2005 model, this tail is larger, with a second peak, near 50$^{\circ}$. 
AIMD results\cite{MaZhangTuckerman} also find this tail, while the NN potential of Behler and co-workers does not.\cite{BehlerPNAS}

The structure of a water molecule's nearest neighbors can be characterized using the  tetrahedral order parameter, Q, of Errington and Debenedetti\cite{ErringtonDebenedetti}
\begin{equation}
Q_i=1-{3 \over 8} \sum_{j=1}^3 \sum_{k=j+1}^4 \left( \cos \theta_{ijk} + {1 \over 3} \right)^2
\label{eq:qtetra}
\end{equation}
where the sums are over the four nearest neighbors of molecule $i$ and $\theta_{ijk}$ is the angle between the three molecules, with molecule $i$ at the center.
This parameter is zero for a random arrangement of neighbors, as  in an ideal gas, and one for a perfectly tetrahedral structure, as in ice Ih.
The distribution of Q is shown in Figure \ref{fig:pbeta}(B), with  the TIP5P/2005 results for comparison.
The average value of Q is 0.66$\pm$0.02 (Table \ref{tab:structure}).
The average Q combining x-ray and neutron diffraction data with empirical potential structural refinement (EPSR) simulations, give a 
value of 0.576.\cite{SoperBenmore} 
(This also depends on isotope, for heavy water Q equals 0.593.\cite{SoperBenmore}, so NQEs appear to make liquid water less tetrahedral.) 
AIMD results give Q values that are more tetrahedral, with Q values ranging from 0.7 to 0.9.\cite{Chen10846,WuAIMD2018} 

The underestimation of the pressure indicates that the density at a pressure
of 1 atm would be too high, as would be found using additional simulations in the 
constant  particle number, pressure,  and temperature (NPT) ensemble. 
Simulations at other temperatures would determine the density as a function of temperature and find if the model gives
a correct temperature of maximum density.
However, given the incorrect pressure at 300 K, we feel that additional simulations would be better to do after the potential 
has been revised to more effectively capture short-range repulsive interactions.

 \vspace{10. pt}

 \noindent
 {\bf Methane in water.}  The water-methane dimer energy is shown in Figure \ref{fig:dimerMethane}. The agreement with the
 {\it ab initio} results\cite{Metz} are very good. 
 The properties of the methane water solution are shown in Figure \ref{fig:methane}. 
 \begin{figure}[h]
    \centering 
        \includegraphics[bb=0.0cm 0.0cm 7.08cm 7.1cm]{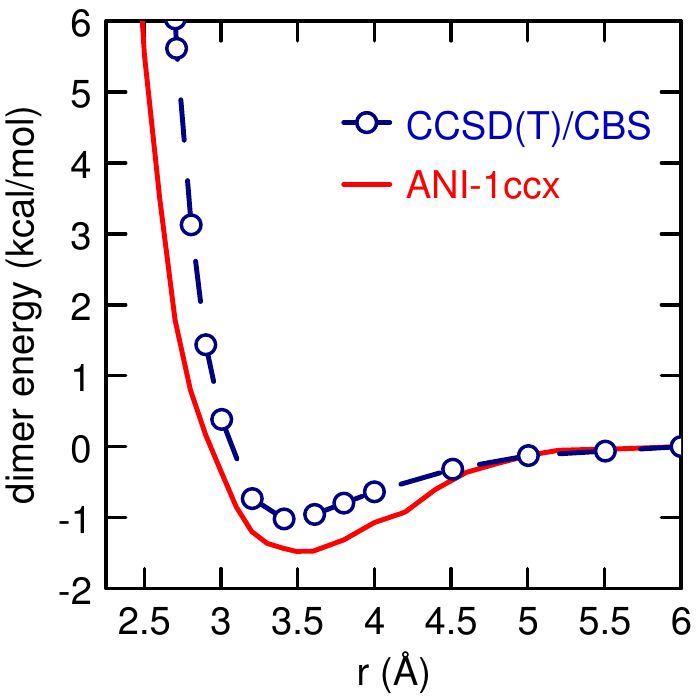}
        \caption{Minimized energy of the water-methane pair as a function of oxygen-carbon separation comparing the ANI-1ccx model to {\it ab initio} results\cite{Metz}.}.
                 \label{fig:dimerMethane}
\end{figure}
 Integrating the oxygen-carbon RDF over this first solvation shell, out to 5.4 {\AA}, gives 20 nearest neighbors.
 Simulation studies find coordinate numbers of 19 to 23,\cite{LueBlankschtein,BridgemanBuckinghamSkipper,MengKollman}
 with AIMD giving 17.\cite{MontagnaSterponeGuidoni} 
 Experimental estimates from neutron diffraction results give 16\cite{KohSoper2000} and 19.\cite{DeJong}
 This value is sensitive to the cut-off distance, the AIMD result of Montagna, {\it et al.},\cite{MontagnaSterponeGuidoni} and the 
 experimental results of Koh, \it et al.},\cite{KohSoper2000} which give lower coordination numbers, use a 5.0 {\AA} limit. The others
 use a limit of around 5.5 {\AA}. Using a 5.0 {\AA} limit with the ANI-1ccx results gives a coordination number of 17. 
  
 From Figure \ref{fig:methane} it is evident that the structure of water in the first solvation shell of methane, from about 3.5 {\AA} to 5.4 {\AA}, 
 is different from the bulk.
 The number of hydrogen bonds increases to over 3.9 and the tetrahedral order increases as well.
 The volume available to a molecule as given by Voronoi volume increases if the standard tessellation is used, but decreases if the S-cell method is used,
 which takes into account different molecular size, as has been seen previously.\cite{IslamFlintRick}
 A number of simulations using AIMD,\cite{MontagnaSterponeGuidoni,Grdadolnik322} all-atom force fields,\cite{Raschke6777,GodecSmithMerzel,Galamba,KimTianWu,AshbaughJPCB2018} and coarse-grained models\cite{IslamFlintRick} 
 and experimental studies\cite{BUCHANAN200589,RezusBakker,Grdadolnik322,DavisDor} have examined the structure of water 
 around hydrophobic molecules.
 The results of these studies are mixed, with some finding a more\cite{Grdadolnik322,DavisDor,RezusBakker,Raschke6777,Galamba,GodecSmithMerzel,MontagnaSterponeGuidoni,AshbaughJPCB2018,IslamFlintRick} and some finding a less\cite{BUCHANAN200589,KimTianWu} structured first solvation shell.
 Our results,  finding an increased number of hydrogen bonds and an increased tetrahedral order, imply more orientational order of water next to 
 methane. 
 A number of simulation studies have found enhanced tetrahedral order in the first solvation shell.\cite{Grdadolnik322,Galamba,AshbaughJPCB2018}
 AIMD\cite{MontagnaSterponeGuidoni,Grdadolnik322} and all-atom potential model simulations\cite{GodecSmithMerzel,AshbaughJPCB2018}
 also find an increased number of hydrogen bonds, while other models find no difference in hydrogen bonds.\cite{Galamba,KimTianWu}

\begin{figure*}
    \centering 
        \includegraphics[bb=0.4cm 0.0cm 7.41cm 7.1cm]{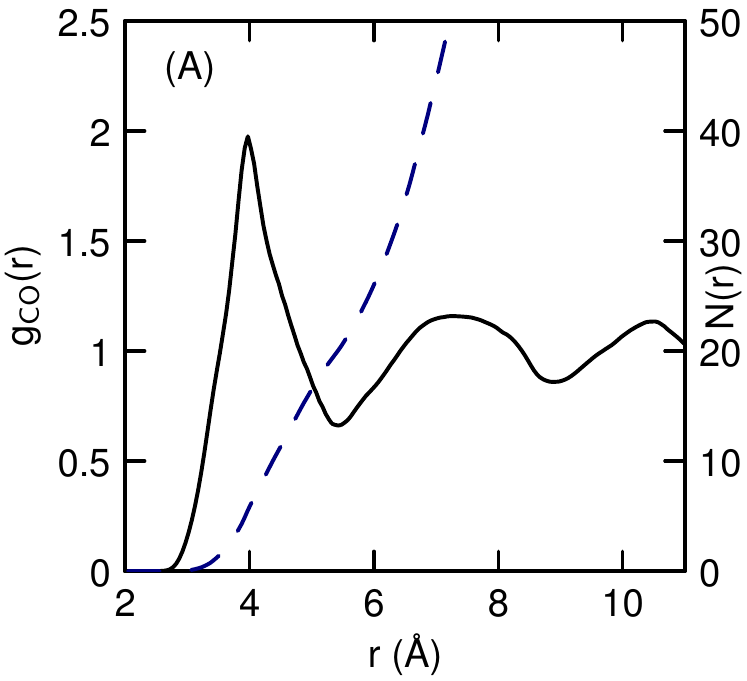}
        \includegraphics[bb=-0.5cm 0.0cm 7.13cm 6.92cm]{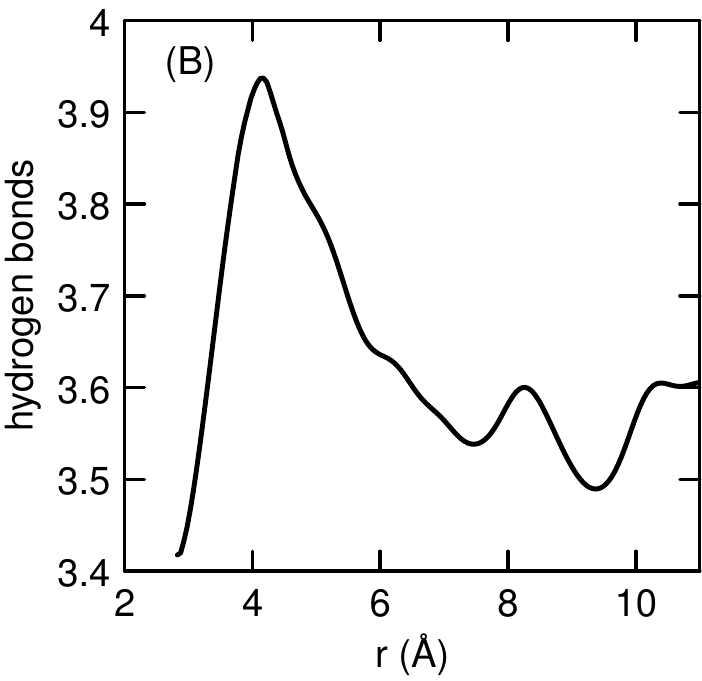}
         \includegraphics[bb=0.65cm 0.0cm 7.17cm 6.92cm]{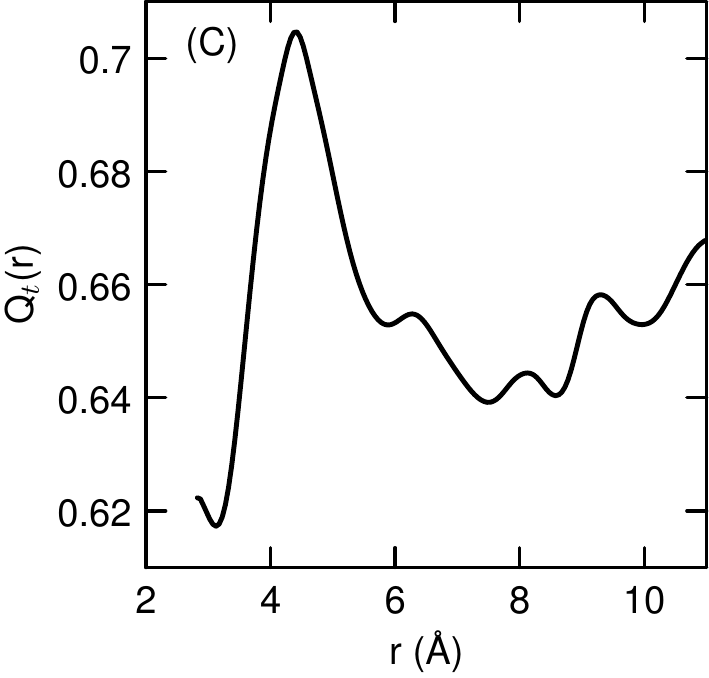}
         \includegraphics[bb=-1.0cm 0.0cm 7.13cm 6.92cm]{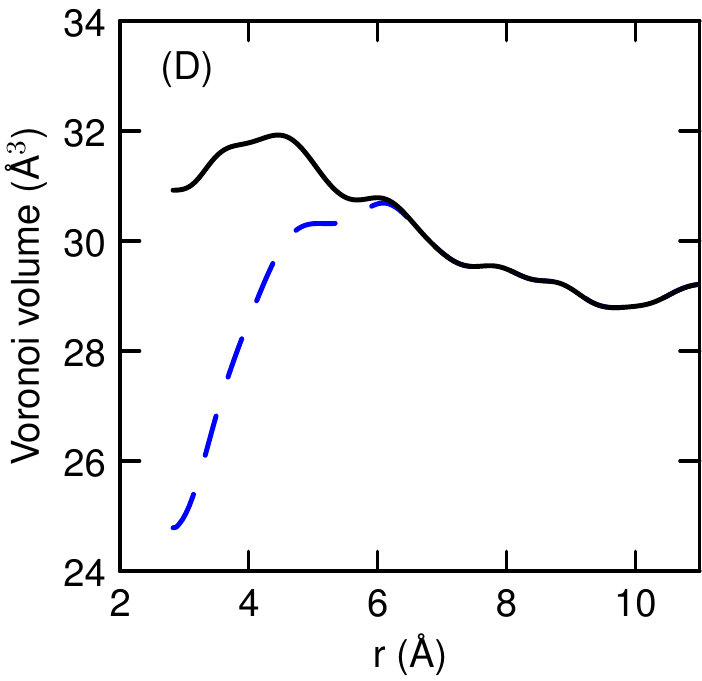}
        \caption{(A) Carbon-oxygen radial distribution function (sold line) and running coordination number, N(r), (blue dashed line) (B) hydrogen bond numbers, (C) tetrahedral order parameter, and (D) standard (solid line) and S-cel (blue dashed line) Voronoi volumes as a function of carbon-oxygen distance. }
\label{fig:methane}
\end{figure*}

\section{Conclusions}

The NI-1ccx potential reproduces the many of the structural properties of liquid water and the methane/water solution, as well as dynamical properties of liquid water, at the standard density and 300 K.
It should be stressed that the model was not developed specifically for water, liquid or otherwise, but as part of a broadly applicable  potential for molecules
containing H,C,N, and O atoms, trained on quantum data for molecules.\cite{SmithRoitberg2018} 
No experimental or theoretical data from the liquid phase was used in the training of the model.
For water, the potential reproduces dynamical properties for both translational and rotational motion (Table \ref{tab:thermo}). The
structure of the liquid as given by the radial distribution functions (Figure \ref{fig:grs}) are in good agreement with experiment,
as good or better than many pair-potentials and AIMD simulations.\cite{CisnerosChemRev,MaZhangTuckerman}
Many-body polarization effects are also described accurately. The hydrogen bonded trimer has hydrogen bonds about 12\% stronger than those
of the dimer, in aggreement of the {\it ab initio} results.\cite{KeutschCruzanSaykally}
Non-additive effects are also evident in the liquid phase energy, as given by $\Delta$H$_{vap}$, compared to the dimer energy. 
The ANI-1ccx potential has the interaction strengths correct for both the dimer and the liquid phase. Non-polarizable potentials, like TIP4P/2005, 
have to overestimate the pair energy to get an accurate $\Delta$H$_{vap}$ (Figure \ref{fig:dimerE} (B)).
For the methane/water solution, the model gives an accurate number of nearest-neighbor water molecules and an increase in the number of hydrogen bonds and tetrahedral
order for the solvating waters, in agreement with experimental and AIMD results.\cite{MontagnaSterponeGuidoni,KohSoper2000,Grdadolnik322}
The one main inaccuracy of the model is the negative pressure for water at the experimental density, indicating the predicted density under standard conditions
would be too high.

The quality of the results is dependent on three factors: the mathematical form, the training, and the execution of the model.
The one major approximation of the form of the potential is that it is short-ranged, going to zero at 
a distance of 5.2 {\AA}. It also ignores polarization beyond the three-body level. 
The training, using single molecules or dimers with energy minimized structures and structures from normal mode sampling to generate structures away from minima, appears to do well near the energy minima but not at close distances (Figures \ref{fig:dimerE} and \ref{fig:dimerMethane}).
For the execution of the model, all atoms were treated
classically, so NQEs are neglected. 
The three factors then lead to three specific aspects of the potential as applied to liquid water: no interactions beyond 5.2 {\AA}, too soft at small distances, and no NQEs.

The influence of NQEs on the properties of water are well-known. As discussed above, treating the hydrogen atoms as quantum particles would 
increase the density (and pressure)\cite{ChengCeriotti,SpureKuhne,FanourgakisXantheas2008} and broaden the widths of the oxygen-hydrogen and hydrogen-hydrogen radial distribution functions, without changing the
oxygen-oxygen very much.\cite{SternBerne,SpureKuhne,ChengCeriotti,MarsalekMarkland}
These would all improve the results of the ANI-1ccx potential, but the effects are not all that large for the density change.
Any deficiencies of the model are not likely to be due to a lack of long-ranged interactions.
For one thing, addition of long-ranged interactions would only make the pressure more negative, correcting it in the wrong direction.
Secondly, the liquid interactions quickly become small above 3.5 {\AA} (Figure \ref{fig:dimerE}(B)), and accurate NN models for water 
have been developed that do not have longed-ranged interactions.\cite{BehlerPNAS}
(Even NN models for protonated water clusters were developed which did not have charges.\cite{C4CP04751F,SchranBehlerMarx})
Long-ranged interactions could be built into a NN model by adding Coulombic interactions with either assigned\cite{WangYang2018} or
NN generated\cite{MorawietzSharmaBehler,GrisafiCeriotti} charges, but this does not appear to be necessary, at least for the systems studied here.
The negative pressure, the most inaccurate result of the ANI-1ccx model for the water simulations, is likely due to the potential being not repulsive enough at
small distances. 
This could be corrected by training with dimer structures at these distances.
The ANI-1ccx potential was trained for dimers only near the energy minimum.\cite{SmithIsayevRoitberg,SmithRoitberg2018}
Alternatively, corrections to the NN potential could be done as the simulation proceeds, on-the-fly, using query by committee methods\cite{SmithRoitberg2018}
or other\cite{DeVita2015} methods to determine when the model may be inaccurate.

The success of the ANI-1ccx potential in simulating a challenging system like water demonstrates the promise of the approach. 
Using carefully selected
data sets of only single molecules and molecule pairs, models can be trained which work well in the liquid phase. 
This enables the study of novel materials for which there may not be experimental data, as well as simulating reacting systems.
The present results indicate that
improved models could be trained by using dimer structures that probe the repulsive part of the potential along  with the potential minimum. 

\section*{Acknowledgements}
This material is based upon work supported by the U.S. Department of Energy, Office of Science, Basic Energy Sciences, under EPSCoR Grant No. DE-SC0012432 with additional support from the Louisiana Board of Regents.

%\section*{References}

%\bibliography{water}

\begin{thebibliography}{10}
\expandafter\ifx\csname url\endcsname\relax
  \def\url#1{\texttt{#1}}\fi
\expandafter\ifx\csname urlprefix\endcsname\relax\def\urlprefix{URL }\fi
\expandafter\ifx\csname href\endcsname\relax
  \def\href#1#2{#2} \def\path#1{#1}\fi

\bibitem{CarParrinello}
R.~Car, M.~Parrinello, Unified approach for molecular dynamics and
  density-functional theory, Phys. Rev. Lett. 55 (1985) 2471--2474.

\bibitem{PaquetViktor}
E.~Paquet, H.~L. Viktor, Computational methods for ab initio molecular
  dynamics, Advances in Chemistry 2018 (2018) 9839641.

\bibitem{Doren1995}
T.~B. Blank, S.~D. Brown, A.~W. Calhoun, D.~J. Doren, Neural network models of
  potential energy surfaces, J. Chem. Phys. 103~(10) (1995) 4129--4137.

\bibitem{Manzhos}
S.~Manzhos, X.~Wang, R.~Dawes, T.~Carrington, A nested molecule-independent
  neural network approach for high-quality potential fits, J. Phys. Chem. A
  110~(16) (2006) 5295--5304.

\bibitem{BehlerParrinello}
J.~Behler, M.~Parrinello, Generalized neural-network representation of
  high-dimensional potential-energy surfaces, Phys. Rev. Lett. 98 (2007)
  146401.

\bibitem{Knoll2015}
O.~A. von Lilienfeld, R.~Ramakrishnan, M.~Rupp, A.~Knoll, Fourier series of
  atomic radial distribution functions: A molecular fingerprint for machine
  learning models of quantum chemical properties, Int. J. Quantum Chem.
  115~(16) (2015) 1084--1093.

\bibitem{DeVita2015}
Z.~Li, J.~R. Kermode, A.~{De Vita}, Molecular dynamics with on-the-fly machine
  learning of quantum-mechanical forces, Phys. Rev. Lett. 114 (2015) 096405.

\bibitem{Behler2016}
J.~Behler, Perspective: Machine learning potentials for atomistic simulations,
  J. Chem. Phys. 145 (2016) 170901.

\bibitem{SmithRoitberg2018}
J.~S. Smith, B.~Nebgen, N.~Lubbers, O.~Isayev, A.~E. Roitberg, Less is more:
  Sampling chemical space with active learning, J. Chem. Phys. 148 (2018)
  241733.

\bibitem{ANI-1CCX}
J.~S. Smith, B.~T. Nebgen, R.~Zubatyuk, N.~Lubbers, C.~Devereux, K.~Barros,
  S.~Tretiak, O.~Isayev, A.~E. Roitberg, Approaching coupled cluster accuracy
  with a general-purpose neural network potential through transfer learning,
  Nat. Commun. 10 (2019) 2903.

\bibitem{Bartok}
A.~P. Bart\'ok, R.~Kondor, G.~Cs\'anyi, On representing chemical environments,
  Phys. Rev. B 87 (2013) 184115.

\bibitem{SmithIsayevRoitberg}
J.~S. Smith, O.~Isayev, A.~E. Roitberg, Ani-1: an extensible neural network
  potential with dft accuracy at force field computational cost, Chem. Sci. 8
  (2017) 3192--3203.

\bibitem{ChmielaTkatchenko}
S.~Chmiela, H.~E. Sauceda, K.~M{\"u}ller, A.~Tkatchenko, Towards exact
  molecular dynamics simulations with machine-learned force fields, Nat.
  Commun. 9 (2018) 3887.

\bibitem{WillattMusilCeriotti}
M.~J. Willatt, F.~Musil, M.~Ceriotti, Atom-density representations for machine
  learning, J. Chem. Phys. 150 (2019) 154110.

\bibitem{Liu2019}
S.~Liu, J.~Li, K.~C. Bennett, B.~Ganoe, T.~Stauch, M.~Head-Gordon, A.~Hexemer,
  D.~Ushizima, T.~Head-Gordon, Multiresolution 3d-densenet for chemical shift
  prediction in nmr crystallography, J. Phys. Chem. Lett. 10~(16) (2019)
  4558--4565.

\bibitem{BehlerReview}
J.~Behler, First principles neural network potentials for reactive simulations
  of large molecular and condensed systems, Angew. Chem. Int. Ed. Engl. 56~(42)
  (2017) 12828--12840.

\bibitem{BehlerPNAS}
T.~Morawietz, A.~Singraber, C.~Dellago, J.~Behler, How van der waals
  interactions determine the unique properties of water, Proc. Natl. Acad. Sci.
  (USA) 113 (2016) 8368--8373.

\bibitem{NatarajanBehler}
S.~K. Natarajan, J.~Behler, Neural network molecular dynamics simulations of
  solid-liquid interfaces: water at low-index copper surfaces, Phys. Chem.
  Chem. Phys. 18 (2016) 28704.

\bibitem{WangYang2018}
H.~Wang, W.~Yang, Force field for water based on neural network, J. Phys. Chem.
  Lett. 9 (2018) 3232--3240.

\bibitem{Singraber}
A.~Singraber, T.~Morawietz, J.~Behler, C.~Dellago, Density anomaly of water at
  negative pressures from first principles, J. Phys.: Condens. Matter 30 (2018)
  254005.

\bibitem{ChengCeriotti}
B.~Cheng, E.~A. Engel, J.~Behler, C.~Dellago, M.~Ceriotti, Ab initio
  thermodynamics of liquid and solid water, Proc. Natl. Acad. Sci. (USA) 116
  (2019) 1110--1115.

\bibitem{MorawietzSharmaBehler}
T.~Morawietz, V.~Sharma, J.~Behler, A neural network potential-energy surface
  for the water dimer based on environment-dependent atomic energies and
  charges, J. Chem. Phys. 136 (2012) 064103.

\bibitem{LiLiPackard}
Y.~Li, H.~Li, F.~C. {Packard IV}, B.~Narayanan, F.~Sen, M.~K.~Y. Chan,
  S.~Sankaranarayanan, B.~R. Brooks, B.~Roux, Machine learning force field
  parameters from ab initio data, J. Chem. Theory Comput. 13 (2017) 4492--4503.

\bibitem{Beraau}
T.~Bereau, R.~A. DiStasio, A.~Tkatchenko, O.~A. von Lilienfeld, Non-covalent
  interactions across organic and biological subsets of chemical space:
  Physics-based potentials parametrized from machine learning, J. Chem. Phys.
  148~(24) (2018) 241706.

\bibitem{NguyenPaesani}
T.~T. Nguyen, E.~Sz\'ekely, G.~Imbalzano, J.~Behler, G.~Cs\'anyi, M.~Ceriotti,
  A.~W. G\"otz, F.~Paesani, Comparison of permutationally invariant
  polynomials, neural networks, and gaussian approximation potentials in
  representing water interactions through many-body expansions, J. Chem. Phys.
  148~(24) (2018) 241725.

\bibitem{TaylorStone}
M.~E. Taylor, P.~Stone, Transfer learning for reinforcement learning domains: a
  survey, J. Mac. Learn. Res. 10 (2009) 1633--1685.

\bibitem{Neese}
C.~Riplinger, P.~Pinski, U.~Becker, E.~F. Valeev, F.~Neese, Sparse maps-a
  systematic infrastructure for reduced-scaling electronic structure methods.
  ii. liner scaling domain based pair natural orbital couple cluster theory, J.
  Chem. Phys. 144 (2016) 024109.

\bibitem{Hjorth_Larsen_2017}
A.~H. Larsen, J.~J. Mortensen, J.~Blomqvist, I.~E. Castelli, R.~Christensen,
  M.~Du{\l}ak, J.~Friis, M.~N. Groves, B.~Hammer, C.~Hargus, E.~D. Hermes,
  P.~C. Jennings, P.~B. Jensen, J.~Kermode, J.~R. Kitchin, E.~L. Kolsbjerg,
  J.~Kubal, K.~Kaasbjerg, S.~Lysgaard, J.~B. Maronsson, T.~Maxson, T.~Olsen,
  L.~Pastewka, A.~Peterson, C.~Rostgaard, J.~Schi{\o}tz, O.~Schütt,
  M.~Strange, K.~S. Thygesen, T.~Vegge, L.~Vilhelmsen, M.~Walter, Z.~Zeng,
  K.~W. Jacobsen, The atomic simulation environment{\textemdash}a python
  library for working with atoms, Journal of Physics: Condensed Matter 29~(27)
  (2017) 273002.

\bibitem{SPCE}
H.~J.~C. Berendsen, J.~R. Grigera, T.~P. Straatsma, The missing term in
  effective pair potentials, J. Phys. Chem. 91 (1987) 6269.

\bibitem{MaZhangTuckerman}
Z.~Ma, Y.~Zhang, M.~E. Tuckerman, Ab initio molecular dynamics study of water
  at constant pressure using converged basis sets and empirical dispersion
  corrections, J. Chem. Phys. 137 (2012) 044506.

\bibitem{Willow2015}
S.~Y. Willow, M.~A. Salim, K.~S. Kim, S.~Hirata, Ab initio molecular dynamics
  of liquid water using embedded-fragment second-order many-body perturbation
  theory towards its accurate property prediction, Sci. Rep. 5 (2015) 14358.

\bibitem{PastanaAIMD2017}
L.~R. Pestana, N.~Mardirossian, M.~{Head-Gordon}, T.~{Head-Gordon}, Ab initio
  molecular dynamics simulations of liquid water using high quality meta-gga
  functionals, Chem. Sci. 8 (2017) 3554.

\bibitem{WuAIMD2018}
L.~Zheng, M.~Chen, Z.~Sun, H.-Y. Ko, B.~Santra, P.~Dhuvad, X.~Wu, Structural,
  electronic, and dynamical properties of liquid water by ab initio molecular
  dynamics based on scan functional within the canonical ensemble, J. Chem.
  Phys. 148~(16) (2018) 164505.

\bibitem{LiuHeZhangQi}
J.~Liu, X.~He, J.~Z.~H. Zhang, L.~W. Qi, Hydrogen-bond structure dynamics in
  bulk water: insights from ab initio simulations with coupled cluster theory,
  Chem. Sci. 9 (2018) 2065--2073.

\bibitem{Chen10846}
M.~Chen, H.-Y. Ko, R.~C. Remsing, M.~F. Calegari~Andrade, B.~Santra, Z.~Sun,
  A.~Selloni, R.~Car, M.~L. Klein, J.~P. Perdew, X.~Wu, Ab initio theory and
  modeling of water, Proc. Natl. Acad. Sci. (USA) 114~(41) (2017) 10846--10851.

\bibitem{YehHummer}
I.~C. Yeh, G.~Hummer, System-size dependence of diffusion coefficients and
  viscosities from molecular dynamics simulations with periodic boundary
  conditions, J. Phys. Chem. B 108 (2004) 15873--15879.

\bibitem{HarrisWoolf}
K.~R. Harris, L.~A. Woolf, J. Chem. Eng. Data 49 (2004) 1064.

\bibitem{Impey}
R.~W. Impey, P.~A. Madden, I.~R. McDonald, Spectroscopic and transport
  properties of water. model calculations and the interpretation of
  experimental results., Mol. Phys. 46 (1982) 513--539.

\bibitem{AnishchikMedvedev}
S.~V. Anishchik, N.~N. Medvedev, Three-dimensional apollonian packing as a
  model for dense granular systems, Phys. Rev. Lett. 75 (1995) 4314--4317.

\bibitem{VoloshinGeiger}
V.~P. Voloshin, N.~N. Medvedev, M.~N. Andrews, R.~R. Burri, R.~Winter,
  A.~Geiger, Volumetric properties of hydrated peptides: Voronoi–delaunay
  analysis of molecular simulation runs, J. Phys. Chem. B 115~(48) (2011)
  14217--14228.

\bibitem{Franks}
F.~Franks, in: F.~Franks (Ed.), Water-A Comprehensive Treatise, Plenum, New
  York, 1972, pp. 1--20.

\bibitem{Metz}
M.~P. Metz, K.~Szalewicz, J.~Sarka, R.~T\'{o}bi\'{a}s, A.~G. Cs\'{a}sz\'{a}r,
  E.~M\'{a}tyus, Molecular dimers of methane clathrates: ab initio potential
  energy surfaces and variational vibrational states, Phys. Chem. Chem. Phys.
  21 (2019) 13504.

\bibitem{GuidezGordon}
E.~B. Guidez, M.~S. Gordon, Dispersion correction derived from first principles
  for density functional theory and hartree-fock theory, J. Phys. Chem. A 119
  (2015) 2161--2168.

\bibitem{AbascalVega}
J.~L.~F. Abascal, C.~Vega, A general purpose model for the condensed phases of
  water: TIP4p/2005, J. Chem. Phys. 123 (2005) 234505.

\bibitem{KeutschCruzanSaykally}
F.~N. Keutsch, J.~D. Cruzan, R.~J. Saykally, The water trimer, Chem. Rev. 103
  (2003) 2533--2577.

\bibitem{Miliordos}
E.~Miliordos, E.~Apr\'a, S.~S. Xantheas, Optimal geometries and harmonic
  vibrational frequencies of the global minima of water clusters (h2o)n, n =
  2–6, and several hexamer local minima at the ccsd(t) level of theory, J.
  Chem. Phys. 139~(11) (2013) 114302.

\bibitem{ClusterTIP4P}
B.~S. Gonz\'alez, E.~G. Noya, C.~Vega, L.~M. Ses\'e, Nuclear quantum effects in
  water clusters: The role of molecular flexibility, J. Phys. Chem. B 114
  (2010) 2484--2492.

\bibitem{JamesWales}
T.~James, D.~J. Wales, J.~{Hern\'andez-Rojas}, Global minima of water clusters,
  [h2o]n, n$<$21, described by a five-site empirical potential, Chem. Phys.
  Lett. 415 (2005) 302--307.

\bibitem{Dorsey}
N.~E. Dorsey, Properties of Ordinary Water-Substance in all its Phases: Water
  Vapor, Water, and all the Ices, {Reinhold Publishing}, New York, 1940.

\bibitem{Schmidt}
J.~Schmidt, J.~VandeVondele, I.-F.~W. Kuo, D.~Sebastiani, J.~I. Siepmann,
  J.~Hutter, C.~J. Mundy, Isobaric-isothermal molecular dynamics simulations
  utilizing density functional theory: An assessment of the structure and
  density of water at near-ambient conditions, J. Phys. Chem. B 113~(35) (2009)
  11959--11964.

\bibitem{GaidukGalli}
A.~P. Gaiduk, F.~Gygi, G.~Galli, Density and compressibility of liquid water
  and ice from first-principles simulations with hybrid functionals, J. Phys.
  Chem. Lett. 6~(15) (2015) 2902--2908.

\bibitem{Willow2016}
S.~Y. Willow, X.~C. Zeng, S.~S. Xantheas, K.~S. Kim, S.~Hirata, Why is
  mp2-water “cooler” and “denser” than dft-water?, J. Phys. Chem. Lett.
  7~(4) (2016) 680--684.

\bibitem{Krynicki}
K.~Krynicki, C.~D. Green, D.~W. Sawyer, Pressure and temperature dependence of
  self-diffusion in water, Discuss. Faraday Soc. 66 (1978) 199--208.

\bibitem{Jonas}
J.~Jonas, T.~{DeFries}, D.~J. Wilbur, Molecular motions in compressed liquid
  water, J. Chem. Phys. 65 (1976) 582--588.

\bibitem{GrisafiPRL}
A.~Grisafi, D.~M. Wilkins, G.~Cs\'anyi, M.~Ceriotti, Symmetry-adapted machine
  learning for tensorial properties of atomistic systems, Phys. Rev. Lett. 120
  (2018) 036002.

\bibitem{GrisafiCeriotti}
A.~Grisafi, M.~Ceriotti, Incorporating long-range physics in atomic-scale
  machine learning, J. Chem. Phys. 151 (2019) 204105.

\bibitem{Skinnerrdf2014}
L.~B. Skinner, C.~J. Benmore, J.~C. Neuefeind, J.~B. Parise, The structure of
  water around the compressibility minumum, J. Chem. Phys. 141 (2014) 214507.

\bibitem{Soper2000}
A.~K. Soper, The radial distribution functions of water and ice from 220 to 673
  k and at pressures up to 400 mpa, Chem. Phys. 258 (2000) 121--137.

\bibitem{Forster-Tonigold}
K.~{Forster-Tonigold}, A.~Gro{\ss}, Dispersion corrected rpbe studies of liquid
  water, J. Chem. Phys. 141~(6) (2014) 064501.

\bibitem{Soper2013}
A.~K. Soper, The radial distribution functions of water as derived from
  radiation total scattering experiments: Is there anything we can say for
  sure?, ISRN Phys. Chem. 2013 (2013) 279463.

\bibitem{Stefan}
S.~Herrig, M.~Thol, A.~H. Harvey, E.~W. Lemmon, A reference equation of state
  for heavy water, J. Phys. Chem. Ref. Data 47~(4) (2018) 043102.

\bibitem{SternBerne}
H.~A. Stern, B.~J. Berne, Quantum effects in liquid water: Path-integral
  simulations of a flexible and polarizable ab initio model, J. Chem. Phys.
  115~(16) (2001) 7622--7628.

\bibitem{SpureKuhne}
T.~Spura, C.~John, S.~Habershon, T.~D. K\"{u}hne, Nuclear quantum effects in
  liquid water from path-integral simulations using an ab initio force-matching
  approach, Molecular Physics 113~(8) (2015) 808--822.

\bibitem{MarsalekMarkland}
O.~Marsalek, T.~E. Markland, Quantum dynamics and spectroscopy of ab initio
  liquid water: The interplay of nuclear and electronic quantum effects, J.
  Phys. Chem. Lett. 8 (2017) 1545--1551.

\bibitem{SoperBenmore}
A.~K. Soper, C.~J. Benmore, Quantum differences between heavy and light water,
  Phys. Rev. Lett. 101 (2008) 065502.

\bibitem{FanourgakisXantheas2008}
G.~S. Fanourgakis, S.~S. Xantheas, Development of transferable interaction
  potentials for water. v. extension of the flexible, polarizable, thole-type
  model potential (ttm3-f, v. 3.0) to describe the vibrational spectra of water
  clusters and liquid water, J. Chem. Phys. 128 (2008) 074506.

\bibitem{FanourgakisXantheas2006}
G.~S. Fanourgakis, S.~S. Xantheas, The flexible, polarizable, thole-type
  interaction potential for water (ttm2-f) revisited, J. Phys. Chem. A 110~(11)
  (2006) 4100--4106.

\bibitem{PaesaniIuchiVoth}
F.~Paesani, S.~Iuchi, G.~A. Voth, Quantum effects in liquid water from an ab
  initio-based polarizable force field, J. Chem. Phys. 127 (2007) 074506.

\bibitem{NAKAMURA1995127}
M.~Nakamura, K.~Tamura, S.~Murakami, Isotope effects on thermodynamic
  properties: mixtures of x(d2o or h2o) + (1 - x)ch3cn at 298.15 k,
  Thermochimica Acta 253 (1995) 127 -- 136.

\bibitem{Kell}
G.~S. Kell, Density, thermal expansivity, and compressibility of liquid water
  from 0$^{\circ}$ to 150$^{\circ}$c: Correlations and table for atmospheric
  pressure and saturation reviewed and expressed on 1968 temperature scale, J.
  Chem. Eng. Data 20 (1975) 97--105.

\bibitem{LuzarChandler}
A.~Luzar, D.~Chandler, Hydrogen-bond kinetic in liquid water, Nature 379 (1996)
  55--57.

\bibitem{SoperBruniRicci}
A.~K. Soper, F.~Bruni, M.~A. Ricci, Site-site pair correlation functions of
  water from 25 to 400$^{\circ}$c: Revised analysis of new and old diffraction data,
  J. Chem. Phys. 106~(1) (1997) 247--254.

\bibitem{PhysRevLett.90.075502}
K.~Modig, B.~G. Pfrommer, B.~Halle, Temperature-dependent hydrogen-bond
  geometry in liquid water, Phys. Rev. Lett. 90 (2003) 075502.

\bibitem{ErringtonDebenedetti}
J.~R. Errington, P.~G. Debenedetti, Relationship between structural order and
  the anomalies of liquid water, Nature 409 (2001) 318--321.

\bibitem{LueBlankschtein}
L.~Lue, D.~Blankschtein, Liquid-state theory of hydrocarbon-water systems:
  application to methane, ethane, and propane, J. Chem. Phys. 96 (1992) 8582.

\bibitem{BridgemanBuckinghamSkipper}
C.~H. Bridgeman, A.~D. Buckingham, N.~T. Skipper, Temperature dependence of
  solvent structure around a hydrophobic solute: a monte carlo study of methane
  in water, Chem. Phys. Lett. 253 (1996) 209--215.

\bibitem{MengKollman}
E.~C. Meng, P.~A. Kollman, Molecular dynamics studies of the properties of
  water around simple organic solutes, J. Phys. Chem. 100 (1996) 11460.

\bibitem{MontagnaSterponeGuidoni}
M.~Montagna, F.~Sterpone, L.~Guidoni, Structural and spectroscopic properties
  of water around small hydrophobic solutes, J. Phys. Chem. B 116 (2012)
  11695--11700.

\bibitem{KohSoper2000}
C.~A. Koh, R.~P. Wisbey, X.~Wu, R.~E. Westacott, A.~K. Soper, Water ordering
  around methane during hydrate formation, J. Chem. Phys. 113 (2000) 6390.

\bibitem{DeJong}
P.~H.~K. {de Jong}, J.~E. Wilson, G.~W. Neilson, A.~D. Buckingham, Hydrophobic
  hydration of methane, Mol. Phys. 91~(1) (1997) 99--104.

\bibitem{IslamFlintRick}
N.~Islam, M.~Flint, S.~W. Rick, Water hydrogen degrees of freedom and the
  hydrophobic effect, J. Chem. Phys. 150 (2019) 014502.

\bibitem{Grdadolnik322}
J.~Grdadolnik, F.~Merzel, F.~Avbelj, Origin of hydrophobicity and enhanced
  water hydrogen bond strength near purely hydrophobic solutes, Proc. Natl.
  Acad. Sci. (USA) 114~(2) (2017) 322--327.

\bibitem{Raschke6777}
T.~M. Raschke, M.~Levitt, Nonpolar solutes enhance water structure within
  hydration shells while reducing interactions between them, Proc. Natl. Acad.
  Sci. (USA) 102~(19) (2005) 6777--6782.

\bibitem{GodecSmithMerzel}
A.~Godec, J.~C. Smith, F.~Merzel, Increase of both order and disorder in the
  first hydration shell with increasing solute polarity, Phys. Rev. Lett. 107
  (2011) 267801.

\bibitem{Galamba}
N.~Galamba, Water’s structure around hydrophobic solutes and the iceberg
  model, J. Phys. Chem. B 117~(7) (2013) 2153--2159.

\bibitem{KimTianWu}
J.~Kim, Y.~Tian, J.~Wu, Thermodynamic and structural evidence for reduced
  hydrogen bonding among water molecules near small hydrophobic solutes, J.
  Phys. Chem. B 119~(36) (2015) 12108--12116.

\bibitem{AshbaughJPCB2018}
H.~S. Ashbaugh, J.~W. Barnett, A.~Saltzman, M.~Langrehr, H.~Houser, Connections
  between the anomalous volumetric properties of alcohols in aqueous solution
  and the volume of hydrophobic association, J. Phys. Chem. B 122~(13) (2018)
  3242--3250.

\bibitem{BUCHANAN200589}
P.~Buchanan, N.~Aldiwan, A.~Soper, J.~Creek, C.~Koh, Decreased structure on
  dissolving methane in water, Chem. Phys. Lett. 415~(1) (2005) 89 -- 93.

\bibitem{RezusBakker}
Y.~L.~A. Rezus, H.~J. Bakker, Observation of immobilized water molecules around
  hydrophobic groups, Phys. Rev. Lett. 99 (2007) 148301.

\bibitem{DavisDor}
J.~G. Davis, K.~P. Gierszal, P.~Wang, D.~Ben-Amotz, Water structural
  transformation at molecular hydrophobic interfaces, Nature 491 (2012)
  582--585.

\bibitem{CisnerosChemRev}
G.~A. Cisneros, K.~T. Wikfeldt, L.~Ojamäe, J.~Lu, Y.~Xu, H.~Torabifard, A.~P.
  Bartók, G.~Csányi, V.~Molinero, F.~Paesani, Modeling molecular interactions
  in water: From pairwise to many-body potential energy functions, Chemical
  Reviews 116~(13) (2016) 7501--7528.

\bibitem{C4CP04751F}
S.~Kondati~Natarajan, T.~Morawietz, J.~Behler, Representing the
  potential-energy surface of protonated water clusters by high-dimensional
  neural network potentials, Phys. Chem. Chem. Phys. 17 (2015) 8356--8371.

\bibitem{SchranBehlerMarx}
C.~Schran, J.~Behler, D.~Marx, Automated fitting of neural network potentials
  at coupled cluster accuracy: Protonated water clusters as testing ground, J.
  Chem. Theory Comput. 16 (2020) 88--99.

\end{thebibliography}

\end{document}